\documentclass[11pt]{article}

\usepackage[margin=1in]{geometry}
\usepackage[T1]{fontenc}
\usepackage[utf8]{inputenc}
\usepackage{lmodern}
\usepackage{microtype}
\usepackage{graphicx}
\usepackage{xspace}
\usepackage[table]{xcolor}
\usepackage{booktabs}
\usepackage{amsmath}
\usepackage{amssymb}
\usepackage{url}
\usepackage[hidelinks]{hyperref}

\providecommand{\Description}[1]{}
\title{NEXT: Reasoning-Driven Video Recommendation via a Vision-Language Model}

\author{%
\small
\begin{tabular}{c}
Yuming Liu$^{1,*}$, Hongye Yang$^{2,*}$, Harrison Zhao$^{2}$, Ellie Zhu$^{3}$\\
Bokai Cao$^{1}$, Lei Huang$^{1}$, Lizhu Zhang$^{1}$, Xiangjun Fan$^{1}$\\[0.4em]
$^{1}$Meta Platforms, Inc., Menlo Park, CA, USA\\
$^{2}$Meta Platforms, Inc., Bellevue, WA, USA\\
$^{3}$Meta Platforms, Inc., New York, NY, USA\\
$^{*}$Equal contribution\\
\scriptsize\texttt{\{yumliu,yhy,harrisonzhao,elliezhu\}@meta.com}\\
\scriptsize\texttt{\{caobokai,leihuang,lizhu,maxfan\}@meta.com}
\end{tabular}
}

\date{}

\begin{document}

\maketitle

\begin{abstract}
We present NEXT (Next-interest EXploration Transformer), a reasoning-driven video recommendation framework that reasons over the video a user has just watched, infers the viewer's next intent, and retrieves concrete follow-up videos. Explicit continuations such as episodes are linked directly; implicit cases are handled by generating intent queries and searching for matching candidates. This Item-to-Intent-to-Item formulation produces directed recommendations beyond co-engagement correlation or semantic similarity.

To make this framework reliable at scale, we train NEXT-8B, a purpose-trained 8B vision-language model with a three-stage recipe: Perception-Enhanced Reinforcement Learning for query-agnostic evidence extraction, Distribution-Aligned Supervised Fine-Tuning over real and synthetic visual QA mixtures, and Group Relative Policy Optimization for last-mile alignment. NEXT-8B achieves the best single-model DocVQA performance, ranking second overall only behind a multi-agent system while surpassing a substantially larger 200B+ scale model, and improves next-intent logic-wise quality by 3.3\% over the base model in a task-specific LLM-as-a-judge evaluation.

We deploy NEXT as an additional retrieval path in a large-scale social media recommendation system and observe statistically significant production gains, including +0.53\% watch time and +0.51\% distinct video exposure. Overall, NEXT shows that a carefully trained compact vision-language model can serve as a practical reasoning engine for next-interest exploration at production scale.

\end{abstract}

\section{Introduction}
Short-video recommenders are strong at fitting engagement logs, but they often miss what should logically come next. This creates three user-facing gaps: popularity bias, reduced content diversity, and broken narrative continuity. A viewer who watches ``Part 1'' may not see ``Part 2''; a viewer who sees an unresolved question may not receive an explanation. These failures come from relying mainly on correlation and semantic similarity rather than reasoning over the current viewing context.

NEXT reframes the recommendation unit from Item-to-Item similarity to Item-to-Intent-to-Item reasoning. Given the video a user has just watched, NEXT first infers a next intent, then retrieves videos satisfying that intent:
\begin{equation}
\begin{aligned}
z_{\text{next}} &= \text{Intent}(v_{\text{current}}, \text{Context}),\\
\hat{v}_{\text{next}} &= \arg\max_{v \in \mathcal{V}} P(v \mid z_{\text{next}}).
\end{aligned}
\end{equation}
For explicit cases such as episode markers or creator continuations, NEXT directly links the next clip. For implicit cases, NEXT generates an intent query such as an explanation, outcome, entity detail, or continuation need, and searches for matching videos.

The main technical challenge is training a compact vision-language model that can reliably extract visual evidence, generate specific next intents, and verify retrieved candidates at production scale. We address this with NEXT-8B, an 8B VLM trained with Perception-Enhanced RL, Distribution-Aligned SFT, and GRPO. NEXT-8B achieves the best single-model DocVQA result, ranking second overall only behind a multi-agent system and outperforming a much larger 200B+ scale model, while providing the reasoning engine for NEXT mining.

Our contributions are: (1) \textbf{NEXT}, an intent-centric video recommendation framework that turns watched videos into explicit continuation links or searchable next-intent queries; (2) \textbf{NEXT-8B}, a compact reasoning VLM whose targeted post-training achieves the best single-model DocVQA performance and improves video next-intent quality; (3) a \textbf{production retrieval path} that integrates high-precision directed NKG edges into an existing recommender; and (4) \textbf{large-scale validation} through offline VLM benchmarks, next-intent LLM judging, and online A/B tests with +0.53\% watch time and +0.51\% distinct video exposure.

\section{Related Work}
\textbf{Recommendation models.} Collaborative filtering~\cite{koren2009matrix}, two-tower retrieval~\cite{yi2019sampling}, DLRM~\cite{naumov2019deep}, and Swing~\cite{yang2020swing} capture co-engagement or item similarity, but they do not explicitly model what a viewer may want next after a specific video. Sequential recommenders model user histories~\cite{kang2018self,sun2019bert4rec}, but they do not directly produce transferable item-level follow-up edges such as ``Episode 1 $\rightarrow$ Episode 2''. Causal recommendation work focuses mainly on debiasing prediction pipelines~\cite{wang2021deconfounded,zhang2021causal}, while NEXT constructs directed transition edges for retrieval.

\textbf{LLMs and VLMs for recommendation.} Recent LLM-based interest exploration maps user behavior to topic transitions~\cite{wang2024llm_exploration}. NEXT operates at a finer granularity: it reasons over multimodal video content, generates a concrete next-intent query, and retrieves videos satisfying that intent. General VLMs such as CLIP~\cite{radford2021learning}, LLaVA~\cite{liu2024visual}, and GPT-4V~\cite{openai2023gpt4v} provide broad multimodal understanding, but are not optimized for query-agnostic evidence extraction and production-scale transition reasoning. NEXT-8B is trained specifically for this role.

\section{NEXT Framework and Model Training}

The NEXT pipeline constructs a \textbf{NEXT Knowledge Graph (NKG)} of directed edges $\langle v_{\text{trigger}}, v_{\text{next}} \rangle$. Unlike symmetric Item-to-Item graphs, NKG is built to answer a next-interest product question: \emph{because the user watched $A$, what logically connected video $B$ may they want next?} The edge is therefore not a similarity shortcut or popularity prior; it is retained only when multimodal evidence, user signals, and NEXT-8B verification support a logical transition from $A$ to $B$.

\paragraph{Design Principles.}
NEXT follows three principles: (1) \textbf{intent generation before retrieval}, so the system searches for what the viewer likely wants next rather than matching by similarity; (2) \textbf{precision-first mining}, so sources are selected for deterministic or logically explicit user signals rather than broad coverage; and (3) \textbf{VLM-based verification}, so candidate edges pass through NEXT-8B before the best one or a small set of grounded results is written to NKG.

Candidate sources are treated as \emph{edge proposers}, not final ranking signals. Explicit continuation edges above threshold receive priority; otherwise NEXT-8B produces a source-agnostic logical verification score, and NKG keeps the top edge or a small verified set using logical fit, source reliability, freshness, and quality filters.

\begin{figure*}[t]
\centering
\includegraphics[width=\textwidth]{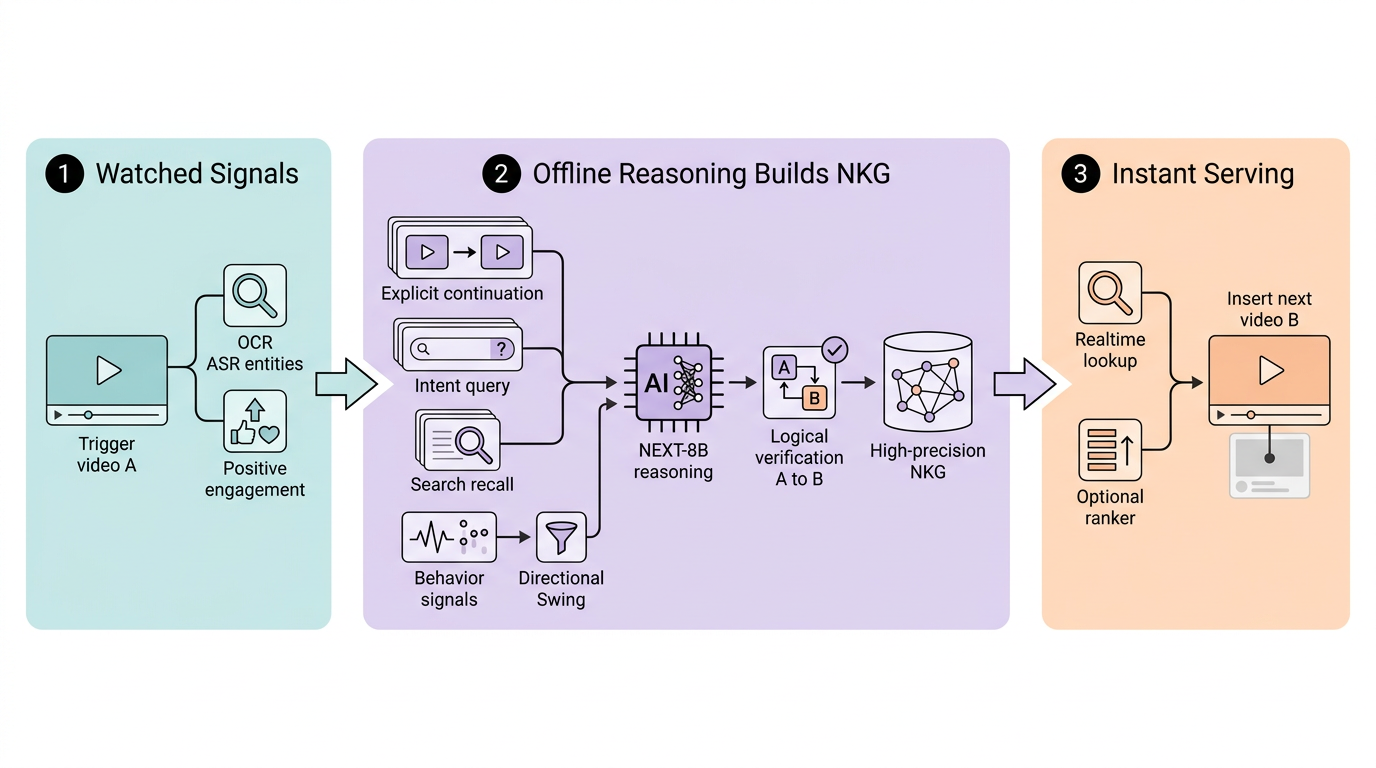}
\caption{NEXT system architecture: watched videos are converted into explicit continuation links or implicit next-intent queries, then retrieved, verified, and served through the NEXT Knowledge Graph.}
\Description{A pipeline diagram showing watched video evidence flowing into explicit continuation detection or implicit NEXT-8B intent generation, then search, verification, NKG storage, online lookup, ranking, and injected next-video serving.}
\label{fig:system}
\end{figure*}

The edge lifecycle is: \textit{evidence extraction} $\rightarrow$ \textit{multi-source candidate generation} $\rightarrow$ \textit{NEXT-8B intent reasoning and verification} $\rightarrow$ \textit{edge scoring and NKG write} $\rightarrow$ \textit{online lookup and insertion}. Figure~\ref{fig:system} illustrates this architecture.

\begin{table}[t]
\centering
\caption{NEXT candidate sources and quality controls.}
\label{tab:candidate_sources}
\resizebox{\columnwidth}{!}{%
\begin{tabular}{l l l}
\toprule
\textbf{Source} & \textbf{Candidate Signal} & \textbf{Quality Control} \\
\midrule
Explicit continuation & episode / sequel cues & deterministic confidence \\
VLM intent-search & generated next-intent & NEXT-8B verification \\
Behavior signal & search / profile / comment & time-window intent filters \\
Directional Swing & ordered co-engagement & SID, direction, denoising \\
\bottomrule
\end{tabular}
}
\end{table}

\subsection{Intent-Centric Mining}

Content-based mining has two paths: explicit continuation detection, where the next video can be linked deterministically, and implicit intent mining, where NEXT-8B generates a next-intent query for retrieval.

\paragraph{Explicit continuation detection.}
When the watched video contains explicit continuation evidence, NEXT directly creates a follow-up edge without broad retrieval. We detect sequential cues from OCR, ASR, creator metadata, and phrases such as ``part two,'' ``next clip,'' and ``to be continued.'' OCR and ASR evidence are fused as:
\begin{equation}
S_{\text{series}} = w_{\text{ocr}} \cdot C_{\text{ocr}} + w_{\text{asr}} \cdot C_{\text{asr}}
\end{equation}
where $C_{\text{ocr}}$ and $C_{\text{asr}}$ are confidence scores from each modality. When $S_{\text{series}} > \tau$, NEXT writes a deterministic continuation edge $V_{\text{part}_n} \rightarrow V_{\text{part}_{n+1}}$ with priority over weaker retrieval signals.

\paragraph{Implicit intent mining.}

For videos without explicit continuation markers, NEXT-8B generates a compact next-intent query from multimodal evidence. Typical intents include answers to Q\&A/riddle videos, outcomes of challenges or experiments, details about mentioned entities, and continuations of incomplete stories. The intent query is sent to an existing video search stack for recall. NEXT-8B then compares the trigger evidence, generated intent, and retrieved candidates, retaining only the candidate or small candidate set with the strongest logical fit before writing edges to NKG. Thus the VLM supplies reasoning and verification, while search supplies scalable recall.

This design separates \emph{what to look for} from \emph{where to retrieve it}. The generated intent is not used as a user-visible explanation; it is a retrieval pivot that converts an unresolved viewing need into a searchable query. During verification, NEXT-8B checks whether candidate $B$ answers, continues, or resolves the need induced by trigger $A$. Candidates that are merely topically similar, popular in the same cluster, or visually close without a follow-up relationship are rejected before NKG write.

\subsection{Behavior-Based Signals}

Behavior traces complement content reasoning by revealing explicit user intent after a trigger video. We emphasize high-certainty patterns rather than broad coverage: (1) \textbf{Watch $\rightarrow$ Search}, where a query shortly after a video indicates an information gap; (2) \textbf{Watch $\rightarrow$ Creator Profile $\rightarrow$ Watch}, where users leave the feed to retrieve a sequel or missing context; and (3) \textbf{Watch $\rightarrow$ Comment}, where comments such as ``part 2?'' or ``what happened after?'' state a next-interest pivot. These signals become retrieval pivots or candidate edges only after time-window, semantic, and intent filters.

\subsubsection{Collaborative Signal: Scene-Aware Reasoning Swing}

To add a collaborative signal without falling back to generic similarity, we use a constrained directional Swing variant that only scores ordered co-engagement patterns consistent with next-video transitions.

\paragraph{Reasoning Swing.}
We extend Swing~\cite{yang2020swing} with three constraints: \textit{directionality}, where users must consume $i$ before $j$ within a session window; \textit{semantic partitioning}, where pairs must share a Semantic ID prefix~\cite{rajput2023recommender}; and \textit{denoising}, where low-support pairs are removed before scoring. These constraints make the collaborative path a logical-transition source rather than a popularity or similarity source.

\textbf{Complete Formulation.}
The final Scene-Aware Reasoning Swing score is:
\begin{equation}
\begin{aligned}
S_{\text{NEXT}}(i, j)
&= \mathbb{I}_{\text{denoise}}(i, j)
\sum_{u \in U_{\text{dir}}(i,j)}
\sum_{\substack{u' \in U_{\text{dir}}(i,j) \\ u' \neq u}} \\
&\quad
\frac{w_u w_{u'} \gamma^{|\Delta t_{u,u'}|}}
{\alpha + |I_{\text{SID}}(u) \cap I_{\text{SID}}(u')|}
\end{aligned}
\end{equation}
where $U_{\text{dir}}$ contains users who consumed $i$ before $j$, $I_{\text{SID}}$ restricts overlap to the same semantic partition, $w_u$ penalizes high-activity users, and $\gamma$ decays inconsistent transition timing. Candidate edges from this path are still verified by NEXT-8B before being written to NKG.

\paragraph{The VLM as an Intent Generation Engine.}
Across all sources, NEXT-8B acts as both a \textit{generator} and a \textit{verifier}. It generates intents for implicit cases, evaluates whether candidate $B$ logically follows from trigger $A$, and selects the strongest edge or small edge set for NKG. To make this practical, all costly multimodal reasoning is executed offline or nearline; online serving is reduced to positive-engagement detection, low-latency edge lookup, and insertion. This architecture creates a dependency on a model that is both highly capable at fine-grained visual reasoning and efficient enough for corpus-scale mining, motivating the training recipe in the next section.

\subsection{NEXT-8B Training}

The central technical challenge is training a compact VLM that can inspect a watched video, extract query-agnostic evidence, generate a next-intent query, and verify retrieved candidates. NEXT-8B uses a standard 8B vision-language architecture; the contribution is its post-training recipe, which enables fine-grained visual understanding, intent generation, and transition reasoning while remaining efficient for large-scale offline mining.

In NEXT, videos are converted into sampled frames and multimodal evidence streams, including OCR, ASR, entities, layout, and scene cues. The model therefore does not need to perform online sequential recommendation directly; instead, it must produce faithful visual evidence, turn that evidence into a precise next-intent, and judge whether a retrieved candidate satisfies the intent. The training recipe below is designed around this evidence-to-reasoning interface.

\subsection{Perception-Enhanced RL}

\begin{figure}[t]
\centering
\includegraphics[width=\columnwidth]{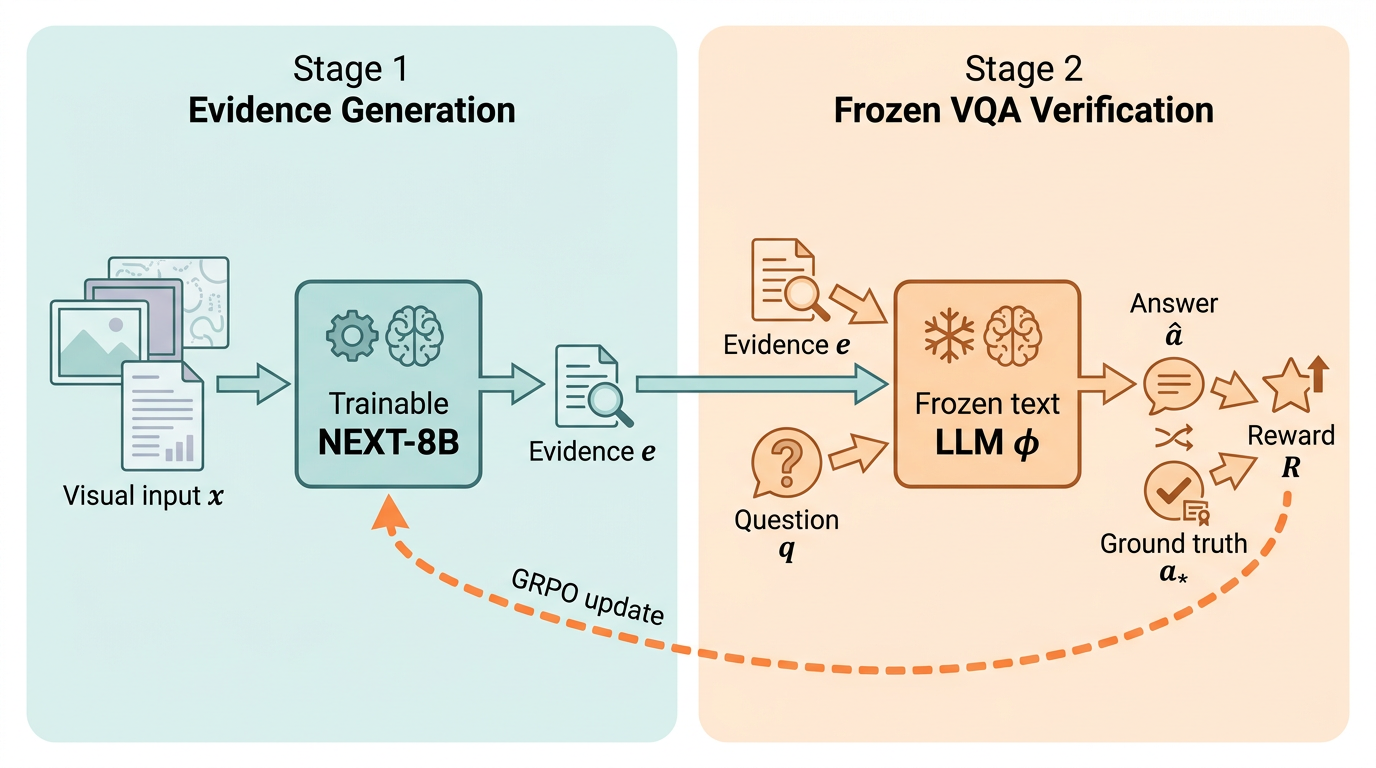}
\caption{Perception-Enhanced RL: Stage 1 trains the multimodal evidence generator to produce query-agnostic evidence optimized for downstream reasoning. Stage 2 uses a frozen text-only LLM whose answer correctness provides the reward signal back to Stage 1.}
\Description{A two-stage training diagram where NEXT-8B generates evidence from visual input, a frozen text verifier answers questions from that evidence, and answer correctness supplies reward back to the policy.}
\label{fig:decoupled_rl}
\end{figure}

General-purpose VLMs often produce fluent summaries while missing small text, entities, spatial relations, or subtle scene cues. Perception-Enhanced RL addresses this by separating \emph{perception/evidence production} from \emph{reasoning/answer production}: a multimodal evidence generator is trained, while a frozen text-only verifier answers questions using only the generated evidence and provides the reward signal (Figure~\ref{fig:decoupled_rl}).

Given visual input $x$, the multimodal policy $\pi_\theta$ generates query-agnostic evidence $e$:
\begin{equation}
e \sim \pi_\theta(\cdot \mid x).
\end{equation}
For each visual input, a frozen text-only verifier $\phi$ answers questions $Q$ using only $e$:
\begin{equation}
\hat{a}_q = \phi(e, q), \quad \forall q \in Q.
\end{equation}
Because $\phi$ is frozen and does not observe $x$, reward improvements must come from better visual evidence:
\begin{equation}
R(e; Q) = \frac{1}{|Q|} \sum_{q \in Q}
r\!\left(\phi(e, q), a_q^{*}\right),
\end{equation}
where $a_q^{*}$ is the ground-truth answer and $r(\cdot,\cdot)$ measures correctness.

\paragraph{Optimization with GRPO.}
We optimize $\pi_\theta$ with Group Relative Policy Optimization (GRPO)~\cite{shao2024deepseekmath}. For each visual input $x_i$, we sample $K$ evidence strings $e_i^{(k)} \sim \pi_{\theta_{\text{old}}}(\cdot \mid x_i)$. Each reward $R_i^{(k)}$ averages verifier correctness over the associated questions. We then compute the group-relative advantage $A_i^{(k)} = R_i^{(k)} - \frac{1}{K}\sum_{k'=1}^{K} R_i^{(k')}$. The policy is updated by minimizing the clipped GRPO objective:
\begin{equation}
\begin{aligned}
\mathcal{L}_{\text{GRPO}} =
&-\sum_{i,k}
\min\!\left(
\rho_i^{(k)} A_i^{(k)},
\operatorname{clip}(\rho_i^{(k)}, 1-\epsilon, 1+\epsilon) A_i^{(k)}
\right) \\
&+ \beta \operatorname{KL}(\pi_\theta \| \pi_{\text{ref}}),
\end{aligned}
\end{equation}
where $\rho_i^{(k)} =
\frac{\pi_\theta(e_i^{(k)} \mid x_i)}
{\pi_{\theta_{\text{old}}}(e_i^{(k)} \mid x_i)}$ is the policy ratio, $\epsilon$ is the clipping range, and the KL divergence term regularizes the policy against a reference policy $\pi_{\text{ref}}$.

\subsection{Distribution-Aligned Augmentation and SFT}

To make NEXT-8B reliable for chart/table/layout reasoning, we combine real and synthetic DocVQA-style data with distribution-aligned supervision. Each iteration samples from:
\begin{equation}
\mathcal{D}^{(t)} = \lambda \, \mathcal{D}_{\text{real}} \;+\; (1-\lambda)\, \mathcal{D}_{\text{syn}}^{(t)},
\end{equation}
For free-form QA with multiple valid strings, we choose the target $y^*$ from acceptable labels $\mathcal{Y}(x)$ that is closest to the model's current prediction $\hat{y}$:
\begin{equation}
y^{*} = \arg\max_{y \in \mathcal{Y}(x)} \text{sim}(\hat{y}, y),
\end{equation}
using normalized Levenshtein similarity:
\begin{equation}
\text{sim}(a, b) = 1 - \frac{d_{\text{Lev}}(a, b)}{\max(|a|, |b|)}.
\end{equation}
The SFT loss is:
\begin{equation}
\mathcal{L}_{\text{DA-SFT}} = -\mathbb{E}_{x \sim \mathcal{D}} [\log \pi_\theta(y^* \mid x)].
\end{equation}
This preserves correctness while reducing label-format mismatch and stabilizing metrics sensitive to minor string errors.

The real pool includes DocVQA-style corpora such as ChartQA~\cite{masry2022chartqa}, TableVQA, and InfoVQA~\cite{mathew2022infographicvqa}; the synthetic pool uses a large VLM teacher and multi-expert consensus labels to cover rare layouts and long-tail question types.

Although these data are not recommendation logs, they stress the same capabilities required by NEXT mining: extracting small but decisive visual details, preserving them as textual evidence, and reasoning from that evidence rather than from generic semantic similarity. This makes DocVQA-style post-training a controlled way to strengthen the model before it is applied to video-derived intent generation and candidate verification.

\subsection{Last-Mile Preference Optimization}

Finally, we apply GRPO for last-mile preference optimization. This stage aligns the model with practical output preferences that are important for downstream use cases: concise answers, stable formatting, faithful use of extracted evidence, and clear boundaries between observed facts and inferred intent. We sample multiple candidate outputs for each prompt, score them with task-specific preference signals, and optimize the same group-relative objective used above. This final stage improves consistency across DocVQA-style answering, next-intent generation, and candidate verification without changing the underlying 8B architecture.


\section{Experiments}

We evaluate NEXT at three levels: DocVQA measures visual evidence extraction, LLM-as-a-judge measures transfer to video next-intent and $A \rightarrow B$ verification, and production A/B tests measure real recommendation impact.

We evaluate NEXT-8B on the official \textbf{DocVQA Task 1} benchmark~\cite{mathew2021docvqa}, a standard testbed for fine-grained visual understanding over text, tables, figures, and layouts. This capability is directly relevant to NEXT because videos are mined through sampled frames and multimodal evidence before graph-level transition reasoning. NEXT-8B achieves the best single-model result on the public DocVQA leaderboard while using only 8B parameters; it ranks second overall only behind a multi-agent system.

\subsection{Experimental Settings}
\label{sec:exp_settings}

\paragraph{Benchmark and Metric.} DocVQA~\cite{mathew2021docvqa} contains over 50,000 questions on 12,767 document images. Results are evaluated on the held-out test set via the official leaderboard using \textbf{Average Normalized Levenshtein Similarity (ANLS)}.

\paragraph{Model Configuration.} NEXT-8B is an 8B VLM using the three post-training stages above. No task-specific architectural modification is used.

\subsection{Performance and Ablation Analysis}
\label{sec:comparison}

We compare NEXT-8B against public DocVQA leaderboard model entries as of January 2026. Table~\ref{tab:docvqa_leaderboard} reports model results and Table~\ref{tab:ablation} ablates the training recipe.

\paragraph{Key Findings.}

\textbf{1. State-of-the-art efficiency.} NEXT-8B reaches \textbf{97.28\%} ANLS, the best single-model DocVQA result and second overall behind a multi-agent system. It surpasses Qwen3-VL, a 235B MoE model, while being nearly \textbf{30$\times$ smaller}, and closes the gap to human performance (98.11\%) to 0.83 ANLS points.

\textbf{2. Perception-heavy gains.} Relative to Qwen3-VL, NEXT-8B is stronger on \emph{Figure/Diagram} (+1.66), \emph{Free-text} (+1.13), and \emph{Image/Photo} (+4.15), matching the goal of Perception-Enhanced RL.

\textbf{3. Complementary training stages.} The ablation shows that Mixed-Domain Data Augmentation gives the largest gain (+0.48), followed by Last-Mile GRPO (+0.32), Distribution-Aligned SFT (+0.23), and Perception-Enhanced RL (+0.15). The Perception-Enhanced RL gain is measured by comparing final models trained with and without that stage, since its benefit is realized through stronger evidence for downstream SFT and GRPO rather than as a standalone QA checkpoint.

\begin{table*}[t]
\centering
\caption{DocVQA test-set comparison among model entries (ANLS). Best results among single-model entries are in \textbf{bold}; the public leaderboard also contains a higher-scoring multi-agent system.}
\label{tab:docvqa_leaderboard}
\resizebox{\textwidth}{!}{%
\begin{tabular}{l c c c c c c c c c}
\toprule
\textbf{Method} & \textbf{Size} & \textbf{ANLS} & \textbf{Fig/Diag} & \textbf{Form} & \textbf{Tab/List} & \textbf{Layout} & \textbf{Free-text} & \textbf{Image} & \textbf{Handwrt.} \\
\midrule
\rowcolor{gray!15} Human Performance & -- & 98.11 & 97.56 & 98.25 & 97.80 & 98.45 & 98.39 & 97.40 & 97.17 \\
\midrule
\textbf{NEXT-8B (Ours)} & \textbf{8B} & \textbf{97.28} & \textbf{95.03} & 98.19 & 97.09 & \textbf{97.68} & \textbf{96.63} & \textbf{96.73} & 95.75 \\
Qwen3-VL~\cite{bai2025qwen3vl} & 235B (MoE) & 97.25 & 93.37 & \textbf{98.37} & \textbf{97.85} & 97.55 & 95.50 & 92.58 & \textbf{96.58} \\
Seed-VL-1.5~\cite{bytedance2025seedvl15} & 20B (act.) & 96.91 & 94.47 & 98.15 & 97.64 & 96.74 & 95.82 & 91.62 & 95.22 \\
Qwen2-VL~\cite{wang2024qwen2vl} & 72B & 96.70 & 92.06 & 98.16 & 97.03 & 96.78 & 96.19 & 91.35 & 94.36 \\
MiMo-VL-7B-RL~\cite{yue2025mimorl} & 7B & 95.01 & 91.59 & 97.12 & 96.58 & 93.89 & 93.44 & 86.01 & 94.58 \\
\bottomrule
\end{tabular}
}
\end{table*}

\begin{table}[t]
\centering
\caption{Ablation study of the NEXT-8B training recipe on the DocVQA test set.}
\label{tab:ablation}
\resizebox{\columnwidth}{!}{%
\begin{tabular}{l c c}
\toprule
\textbf{Configuration} & \textbf{ANLS (\%)} & \textbf{Gain} \\
\midrule
1. Base Model (Qwen3-VL-8B-Instruct~\cite{bai2025qwen3vl}) & 96.10 & -- \\
2. + Mixed-Domain Data Augmentation & 96.58 & +0.48 \\
3. + Distribution-Aligned SFT & 96.81 & +0.23 \\
4. + Perception-Enhanced RL & 96.96 & +0.15 \\
5. + Last-Mile GRPO Alignment (Full Model) & \textbf{97.28} & +0.32 \\
\bottomrule
\end{tabular}
}
\end{table}

\subsection{VLM Quality and Recommendation Edge Quality}
\label{sec:vlm_edge_quality}

DocVQA is a controlled static proxy; NEXT also requires dynamic next-intent reasoning from video-derived inputs. We therefore add a task-specific LLM-as-a-judge evaluation on content-based mining outputs.

\paragraph{Evaluation setup.}
We sample 1,000 videos, run the base VLM and NEXT-8B on the same video-derived inputs, retrieve candidates by searching each generated next-intent, and ask a strong LLM-as-a-judge evaluator to score \textit{logic-wise quality} on a 1--5 scale. Model identity is hidden and output order is randomized; the judge sees the trigger evidence, generated intent, and candidate evidence, and scores grounding, specificity, candidate fit, and overall logic-wise quality. We report the relative lift in high-quality rate; high-quality rate is defined as the fraction of examples receiving a judge score of at least 4.

\begin{table}[t]
\centering
\caption{LLM-as-a-judge evaluation of generated next-intents and retrieved follow-up candidates. We report relative lift in high-quality rate, where high quality means judge score $\geq$4 on a 1--5 scale.}
\label{tab:llm_judge}
\resizebox{\columnwidth}{!}{%
\begin{tabular}{l c}
\toprule
\textbf{Dimension} & \textbf{High-quality Rate Lift} \\
\midrule
Overall logic-wise quality & \textbf{+3.3\%} \\
Evidence grounding & +4.0\% \\
Intent specificity & +0.2\% \\
A-to-B candidate fit & +3.5\% \\
\bottomrule
\end{tabular}
}
\end{table}

\paragraph{Results.}
As summarized in Table~\ref{tab:llm_judge}, NEXT-8B achieves a \textbf{3.3\% relative improvement} in high-quality logic-wise outputs over the base model, showing that the post-training gains transfer from static visual reasoning to dynamic next-intent generation. The largest improvement appears in evidence grounding, where NEXT-8B more consistently uses observed entities, series cues, unresolved questions, and event outcomes. Intent specificity is closer between the models, while NEXT-8B improves A-to-B candidate fit by avoiding generic similar-content intents that do not form a natural next-video transition.

\paragraph{Interpretation.}
The judge results match the intended role of NEXT-8B. The model is not primarily making queries longer or more specific; instead, it better preserves the evidence needed to form a correct follow-up and rejects candidates that only share surface semantics. This is important for NEXT because online insertion assumes that offline NKG edges are already precise enough to be acted on immediately after positive engagement.

\section{Online A/B Test}

We validate NEXT in a production video recommendation system, where it serves as an additional reasoning-driven retrieval path.

\subsection{System Architecture}

As shown in Figure~\ref{fig:system}, NEXT has two production components: offline NKG construction and online next-video injection. The key deployment choice is to move reasoning out of the request path: offline jobs generate, retrieve, and verify high-precision directed edges, while online serving only detects whether the current watched video has triggered positive engagement and immediately activates the corresponding next-video edge.

\paragraph{Offline Transition Generation.} The offline pipeline populates a \textbf{Real-time NEXT DB} with high-precision NKG edges. Explicit continuations, VLM intent-search candidates, behavior-derived pivots, and constrained collaborative candidates are consolidated into a unified edge pool. NEXT-8B verifies whether each candidate $v_{\text{next}}$ logically follows from $v_{\text{trigger}}$ and keeps only the strongest candidate or a small set of top candidates per trigger. Edge writes include source, verification score, and freshness, enabling serving-time ranking without re-running the VLM.

\paragraph{Online Serving and Immediate Sequence Injection.} Online NEXT serving is triggered by positive engagement with a served trigger video $v_1$, such as high completion rate, a like, a save, or other strong feedback. Once this signal is detected, the system queries the Real-time NEXT DB for precomputed next-interest candidates $\{v_{1.a}, v_{1.b}, \ldots\}$. If only one NEXT candidate is available, the system directly inserts $v_{1.a}$ into the user's upcoming feed. If multiple high-precision NEXT candidates are available, the system reuses the existing production online ranking model, restricted to this NEXT-generated candidate set, to select the best candidate for the current user. The selected candidate is then inserted directly. Thus expensive multimodal reasoning remains offline, while the online path is reduced to engagement detection, NKG lookup, optional ranking within NEXT candidates, and immediate next-video insertion.

We intentionally avoid adding a separate NEXT-specific online ranker. Since NKG candidates have already passed offline logical verification, the online model only resolves personalization among a small high-precision set: for example, choosing which verified follow-up best matches the current user's real-time preferences. This keeps the serving integration simple and makes NEXT an additive retrieval path rather than a replacement for the production recommender.

\subsection{A/B Test Results}

We run large-scale live A/B tests for multiple weeks on a commercial short-form video platform at a scale of approximately 100 million users. The control uses an optimized production recommender with multiple retrieval paths, including HSTU-based sequential models~\cite{zhai2024actions}; treatment adds NEXT as an additional content-reasoned transition source. Table~\ref{tab:ab_results} shows statistically significant gains: the 95\% confidence interval excludes zero and the two-sided p-value is below 0.05 for each primary metric. We observe no statistically significant regression in negative feedback, safety, or latency guardrails.

\begin{table}[t]
\centering
\caption{Online A/B test results: lift in key metrics for NEXT vs.\ production baseline.}
\label{tab:ab_results}
\resizebox{\columnwidth}{!}{%
\begin{tabular}{l c c c}
\toprule
\textbf{Metric} & \textbf{Lift} & \textbf{95\% CI} & \textbf{p-value} \\
\midrule
Watch Time & +0.53\% & excludes 0 & $< 0.05$ \\
Distinct Video Exposure & +0.51\% & excludes 0 & $< 0.05$ \\
\bottomrule
\end{tabular}
}
\end{table}

\paragraph{User Satisfaction and Content Discovery.} NEXT improves watch time by \textbf{+0.53\%} and distinct video exposure by \textbf{+0.51\%}. These gains indicate that reasoning-linked recommendations are engaging while exposing users to a broader set of relevant videos beyond popularity-driven retrieval.

\paragraph{Evaluation methodology.} Offline metrics such as Precision, Recall, and NDCG are misaligned with NEXT because they reward replaying historical engagement and penalize novel but logically coherent follow-ups. Live A/B testing is therefore the most faithful evaluation for an exploration-oriented retrieval path.

\section{Conclusion and Future Work}

We presented NEXT, a reasoning-driven video recommendation framework that recommends what should logically come next after a user watches and positively engages with a video. The system comprises two core contributions: (1) an intent-centric mining framework that consolidates explicit continuation cues, video-understanding intents, behavior-derived signals, and constrained collaborative transitions into high-precision NKG edges, and (2) NEXT-8B, a purpose-trained 8B vision-language model whose three-stage post-training recipe---Perception-Enhanced RL, Distribution-Aligned SFT, and Last-Mile GRPO---achieves the best single-model DocVQA result (97.28\% ANLS), ranking second overall only behind a multi-agent system while surpassing a model nearly 30$\times$ larger. NEXT-8B provides the intent-generation and logical-verification engine for NEXT. At serving time, this NKG enables direct next-video insertion once positive engagement is detected; if multiple verified edges exist, the existing online ranking model selects the best NEXT candidate for the current user. We validate NEXT through offline visual-reasoning benchmarks, task-specific LLM-as-a-judge evaluation, and production A/B tests where NEXT improves watch time (+0.53\%) and distinct video exposure (+0.51\%).

\paragraph{Limitations.} Despite these results, applying a large vision-language model to analyze the full video corpus remains computationally expensive. While our offline/nearline architecture avoids placing the VLM on the online critical path, the throughput of corpus-scale transition mining is still a bottleneck, particularly for rapidly growing content libraries. Further improving the inference efficiency of the mining pipeline---through model distillation, speculative decoding, or more aggressive batching strategies---is an important direction for making NEXT more scalable and responsive to emerging content.

\paragraph{Future Work.} We identify three key directions for future research. First, we aim to jointly optimize \textbf{transition quality and user engagement}: the current system mines directed edges offline and ranks only within verified NEXT candidates online, but a tighter integration that accounts for real-time user behavioral signals during edge scoring could improve relevance while preserving logical coherence. Second, we plan to extend the online serving layer with \textbf{multi-step transition reasoning}, enabling the system to chain multiple transitions (e.g., $v_1 \rightarrow v_{1.a} \rightarrow v_{1.a.b}$) that more deeply align with a user's evolving interest trajectory, rather than limiting injection to single-hop lookups. Third, we seek to better \textbf{balance reasoning and diversity}: while reasoning-driven recommendations improve narrative continuity, over-reliance on a single transition chain may narrow the user's exposure; incorporating diversity-aware re-ranking that interleaves NEXT candidates with serendipitous discoveries is essential for maintaining a healthy and engaging content ecosystem.

\section*{GenAI Usage Disclosure}
The authors used GenAI tools to assist with language polishing, proofreading, and formatting adjustments. All GenAI-assisted edits were reviewed and approved by the authors. The research ideas, method design, experiments, data analysis, and reported results were completed by the authors.

\bibliographystyle{plain}
\bibliography{mybib}

\end{document}